\documentclass{article}

\usepackage[dvips]{graphicx}
\usepackage{eusipco}                       

\title{Wavelets on the sphere.\\ Application to the detection problem}

\name { J. L. Sanz$^1$, D. Herranz$^1$, M. L\'opez-Caniego$^{1,2}$,
F. Arg\"ueso$^3$, }

\address{ $^1$ Instituto de F\'\i{sica} de Cantabria (CSIC-UC), 39005,
  Santander, Spain \\ email: sanz@ifca.unican.es\\ $^2$ Departamento de
  F\'\i{sica} Moderna, Universidad de Cantabria, 39005, Santander, Spain
  \\ $^3$ Departamento de Matem\'aticas, Universidad de Oviedo, 33007,
  Oviedo, Spain }

\begin{document}

\maketitle

\begin{abstract}

A new method is presented for the construction of a natural
continuous wavelet transform on the sphere. It incorporates the
analysis and synthesis with the same wavelet and the definition of
translations and dilations on the sphere through the spherical
harmonic coefficients. We construct a couple of wavelets as an
extension of the flat {\it Mexican Hat Wavelet} to the sphere and
we apply them to the detection of sources on the sphere. We remark
that no projections are used with this methodology.

\end{abstract}


\section{Introduction}

Multiscaling analysis techniques dealing with the analysis/synthesis
of nD-images defined on intervals of $R^n$ have been applied in many
fields of physics in the last 15 years. 

For instance, in the case $n = 1$ one has electronics and audio
signals, in the case $n = 2$ one has optical or infrared images
whereas for $n = 3$ one deals with fluid dynamics or the large-scale
structure of the universe as 3D-images. However, there are data given
on other manifolds like the circle $S_1$ (e. g. scanning along circles
the microwave sky) and the sphere $S_2$ (e. g. geophysics). In this
paper, we are interested in data distributed on the sphere. Trivially,
for the study of local properties (e. g. detection of objects) one can
project on the tangent plane at any point on the sphere to make this
type of analysis but when global properties are taken into account the
curvature of the sphere can not be neglected.

A first approach to deal with these global properties is to make some
global projection of all the points of the sphere. The stereographic
projection has been recently used dealing with the continuous wavelet
transform. In this case, to get the wavelet coefficient at any point
on the sphere, one projects from the opposite point to the local
tangent plane. [1] have made a connection to group theory. The
translations and dilations in the wavelet have their definition on the
plane. Clearly, such a projection does not take into account the
topological structure of the sphere. Some applications to cosmology,
in particular the study of anisotropies of the cosmic microwave
background radiation have been done by some authors ([3],[7],[10])
using the projection of the {\it Mexican hat wavelet}. A drawback of
such projection is the obvious deformation of the pixels and wavelets
near the projection pole. We remark that the synthesis can be done in
terms of another biorthogonal wavelet [11].

Another approach uses some analyzing wavelet functions that are
defined in terms of spherical harmonics [5] with a definition of the
dilation operator and conditions on the wavelets in such a way to get
a synthesis formula. The drawback of such methodology is: the
dilations do not satisfy the appropriate flat limit in general. Also
some examples of wavelet functions are poorly localized
(e. g. Abel-Poison wavelets).

A different approach assumes from the beginning discrete wavelets
incorporating tensor product approaches in polar coordinates, then the
two poles are singular points regarding approximation/stability
properties ([4],[6]). Another approach is adapted to arbitrary point
systems or triangulations on the spheres, then there is no efficient
tool as fast wavelet algorithms. In the approach by [8] basis are
defined on a quasi-uniform icosahedral triangulation on the sphere
allowing for a fast algorithm. However, biorthogonal wavelets are
needed and a lifting scheme for the multiresolution is applied
avoiding the concepts of translations and dilations and also it is not
clear whether the construction leads to a stable $L_2 (S_2)$ basis.
Haar-type wavelets have been developed using different pixel
combinations ([2],[7],[9]).  The first case uses the lifting scheme
weighting for the area of the pixels whereas in the other two cases an
equal area pixelization is used but the Haar-type transform is only
applied on regions of the sphere covering only $\frac{1}{12}$ of the
total area. Clearly, with any pixelization the symmetry on the sphere
is lost.

In this paper we will consider a continuous approach, we will introduce
a methodology that incorporates the analysis and synthesis of any
function defined on the sphere $S_2$ using the same
circularly-symmetric wavelet and also we will introduce the
generalization of the translations/dilations. In this sense we follow
Freeden's approach working with spherical harmonics. Examples will be
given that have the appropriate flat limit. Finally, the application to
the detection of a spot is given, studying the concentration of the
wavelet coefficients.

\section{Properties of the wavelet}

We will consider a circularly-symmetric filter defined on the sphere
$ S_2$

\begin{equation}
\label{eq:cc}
 \Psi(\vec{n}\cdot \vec{\gamma}; R),
\end{equation}

\noindent where $\vec{n}$ is a fixed direction. $\vec{\gamma}$ is
another fixed, but arbitrary direction, therefore $\vec{n}\cdot
\vec{\gamma}$ will represent a rotation on the sphere with respect
to the direction $\vec{n}$ defined by the angle $\theta$
($\cos(\theta)\equiv \vec{n}\cdot \vec{\gamma})$. $R>0$ will
represent a dilation, which will be defined later on through the
spherical harmonics.

We assume the following properties of the filter:

(i) the analysis of any function $f(\vec{n})$ will be done with 
    the wavelets $\Psi(\vec{n}\cdot \vec{\gamma}; R)$,

(ii) the synthesis of any function $f(\vec{n})$ will be done with the
     wavelets coefficients and the wavelets $\Psi(\vec{n}\cdot
     \vec{\gamma}; R)$,

(iii) it will incorporate the definition of translation and dilation
on the sphere.

We remark that no assumption about compensation of the filter 
(i. e. $\int d\Omega (\vec{n})\,\Psi(\vec{n}\cdot \vec{\gamma}; R) = 0$) 
and projection from $R^2$ to $S_2$ is imposed.

\section{Analysis with the filter $\Psi$}

We define the wavelet coefficients associated to the translation
$\vec{\gamma}$ and dilation $R$ for the function $f(\vec{n})$ defined
on $S_2$
\begin{equation}
\label{eq:cd}
 w(R, \vec{\gamma} ) = \int d\Omega (\vec{n})\,f(\vec{n})
 \Psi(\vec{n}\cdot \vec{\gamma}; R).
\end{equation}
\noindent Let us assume the standard decomposition of $f(\vec{n})$ in
spherical harmonics $Y_{lm}(\vec{n})$
\begin{equation}
\label{eq:ce}
 f(\vec{n}) = \sum_{lm} f_{lm}Y_{lm}(\vec{n}), \ \ \ 
 f_{lm} = \int d\Omega (\vec{n})\,f(\vec{n})Y_{lm}^*(\vec{n}).
\end{equation}
\noindent By introducing Eq.(\ref{eq:ce}) into Eq.(\ref{eq:cd}) and
taking into account that $Y_{lm}(\vec{n})$ is an orthonormal base of
$S_2$, we obtain
\begin{equation}
\label{eq:cf}
w(R, \vec{\gamma} ) =
\sum_{lm}(\frac{4\pi}{2l+1})f_{lm}\Psi_l(R)Y_{lm}(\vec{\gamma}),
\end{equation}
\noindent where the Legendre coefficients associated to the
circularly-symmetric filter $\Psi$ are given by
\begin{eqnarray}
\Psi(\vec{n}\cdot \vec{\gamma}; R) & = & \sum_l \Psi_l(R)P_l(\vec{n}\cdot
\vec{\gamma}),\nonumber \\
\label{eq:cg}
\Psi_l(R) & = & (l+\frac{1}{2})\int_{-1}^1dy\,P_l(y)\Psi(y; R).
\end{eqnarray}

\section{Synthesis with the filter $\Psi$}

Now, let us show that in order to have a reconstruction equation,
i. e. $f(\vec{n})$ as a functional integral of the wavelet
coefficients and the wavelet base $\Psi$ one can impose the condition
\begin{equation}
\label{eq:ch}
\Psi_l(R) \equiv (\frac{2l+1}{4\pi})\psi (lR),
\end{equation}
\noindent i. e. $\Psi_l(R)$ depends on the product $lR$ and $\psi(l)$
satisfies the admissibility condition 
\begin{equation}
\label{eq:cl}
C_{\psi} \equiv \int_0^{\infty} \frac{dl}{l}\psi^2(l) < \infty,
\end{equation}
\noindent where $l$ runs in the interval $[0, \infty)$. We remark that
the analogous condition to have a reconstruction on the plane by
substituting $l \rightarrow q$, $q$ being the wave number in Fourier
space. Therefore, the filter $\Psi$ -given by Eq. (\ref{eq:cg})-
can be rewritten as
\begin{equation}
\label{eq:ci}
\Psi(\vec{n}\cdot \vec{\gamma}; R) = \sum_l \Psi_l(R)P_l(\vec{n}\cdot
\vec{\gamma}) = \sum_{lm}\psi
(lR)Y_{lm}^*(\vec{n})Y_{lm}(\vec{\gamma}).
\end{equation}
\noindent Firstly, we remark that the previous equation defines a
dilation on the sphere in terms of dilation of the number $l$ and a
translation on the sphere in terms of a rotation through the spherical
harmonics $Y_{lm}(\vec{\gamma})$. We think that such a definition is
the most natural on the sphere and generalizes the one associated to
dilations and translations in the plane $R^2$ via Fourier space.

Secondly, we can write the following equation
\begin{eqnarray}
\int \frac{dR}{R}\int d\Omega (\vec{\gamma})\,w(R, \vec{\gamma)})
\Psi(\vec{n}\cdot \vec{\gamma}; R) = \nonumber \\
\sum_{lm}f_{lm}Y_{lm}(\vec{n})[\int
\frac{dR}{R}{(\frac{4\pi}{2l+1})}^2\Psi_l^2(R)],  \label{eq:cj}
\end{eqnarray}
\noindent where we have taken the harmonic expansions for $w(R,
\vec{\gamma})$ and $\Psi(\vec{n}\cdot \vec{\gamma}; R)$. If one wants
to have this equation proportional to $\sum_{lm}f_{lm}Y_{lm}(\vec{n})
= f(\vec{n})$, i.e. to be able to reconstruct $f(\vec{n})$, then it is
obvious that necessarily
\begin{equation}
\label{eq:ck}
\int \frac{dR}{R}{(\frac{4\pi}{2l+1})}^2\Psi_l^2(R) = C_{\psi},
\end{equation}
\noindent where $C_{\psi} \neq 0$ must be a constant. A particular
solution to the previous equation is given by Eq. \ref{eq:ch} and the
admissibility condition. In this case, the synthesis equation can be
written as
\begin{equation}
\label{eq:cm}
f(\vec{n}) = \frac{1}{C_{\psi}}\int \frac{dR}{R}\int d\Omega
(\vec{\gamma}) \,w(R, \vec{\gamma})\,\Psi(\vec{n}\cdot \vec{\gamma};
R).
\end{equation}
\noindent and the Equation (\ref{eq:cf}) can be rewritten as
\begin{equation}
\label{eq:cn}
w(R, \vec{\gamma} ) = \sum_{lm}f_{lm}\psi (lR)Y_{lm}(\vec{\gamma}).
\end{equation}
\noindent These equations are the analysis/synthesis counterparts on
$S_2$ of the corresponding ones on $R^2$.

\section{Properties of the filter $\Psi$}

Let us focus on some of the properties of the filter:

{\bf $\Psi$ is a {\it compensated} filter}

Taking into account Eq. (\ref{eq:cl}), $C_{\psi} < \infty$ implies that 
$\psi (l)\rightarrow l^{\epsilon},\epsilon >0$ as $l\rightarrow 0$, i. e. 
$\psi (l=0) = 0$. Now, taking into account Eq. (\ref{eq:ci}), one obtains
\begin{equation}
\label{eq:co}
\int d\Omega (\vec{n})\,\Psi(\vec{n}\cdot \vec{\gamma}; R) = \psi (0) = 0,
\end{equation}
\noindent the filter is compensated (hereinafter wavelet).

{\bf Energy of the wavelet}

Taking into account Eq. (\ref{eq:ci}) and the orthonormal property of
the spherical harmonics, one obtains
\begin{equation}
\label{eq:cp}
\int d\Omega (\vec{n})\,\Psi^2(\vec{n}\cdot \vec{\gamma}; R) = 
\sum_l \frac{2l+1}{4\pi}\psi^2(lR).
\end{equation}

{\bf Energy of the function $f(\vec{n})$}

Taking into account the standard properties of the spherical harmonics
\begin{equation}
\label{eq:cq}
\parallel f \parallel ^2 \equiv \int d\Omega (\vec{n})\,f^{2}(\vec{n})
= \sum_{lm}f^2_{lm},\ \ \ f^2_{lm}\equiv f_{lm}f_{lm}^*.
\end{equation}
\noindent We can also prove the following equivalence
\begin{equation}
\label{eq:cr}
\parallel f \parallel ^2 = \frac{1}{C_{\psi}}\int \frac{dR}{R}\int
d\Omega (\vec{n})\,w^2(R, \vec{n}).
\end{equation}

\section{An example}

\begin{figure*}
\begin{center}
\includegraphics[width=0.9\textwidth]{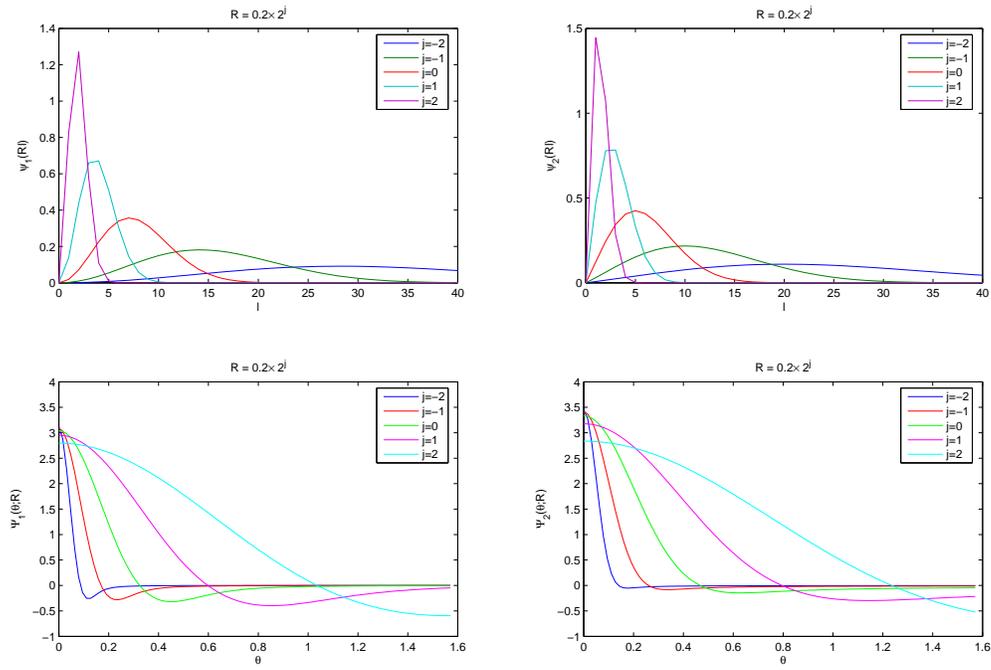} \caption{{\small
Harmonic coefficients of the wavelets $\psi_1$ (top left) and
$\psi_2$ (top right). In the bottom panels, the dilated wavelets
$\psi_1$ (left) and $\psi_2$ (right) are shown on real space. For all
the cases,  $R=0.2\times 2^j$
\label{fig:fig1}}}
\end{center}
\end{figure*}

\begin{figure*}
\begin{center}
\includegraphics[width=0.9\textwidth]{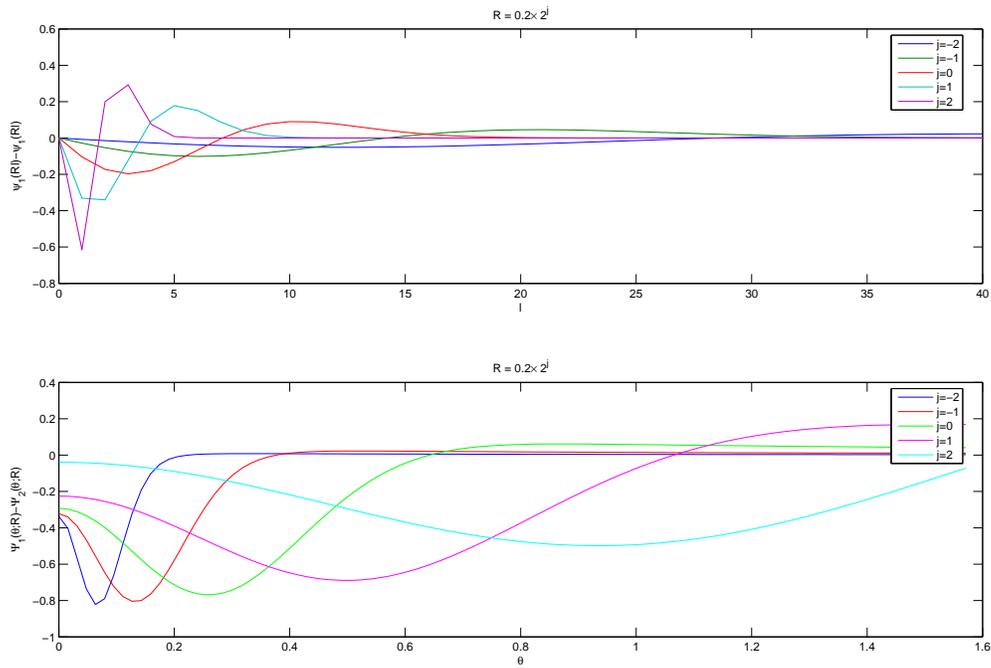} \caption{{\small
Difference $\psi_1 - \psi_2$ on harmonic (top) and real space (bottom)
for the same cases as in Figure~\ref{fig:fig1}
\label{fig:fig2} }}
\end{center}
\end{figure*}

As an example of the previous ideas we will consider the
generalization to the sphere of the {\it Mexican Hat wavelet}
(MHW). We focus on the MHW because it is a widely used tool in
Astronomy, well suited for the detection of pointlike objects such as
extragalactic sources ([3],[7],[10]), but the same ideas can be
applied to other wavelet families as well. Two natural generalizations
of the mother Mexican Hat wavelet on the plane are possible
\begin{equation}
\label{eq:cs}
\psi_1(l)\propto l^2e^{-\frac{1}{2}l^2},\ \ \ 
\psi_2(l)\propto l(l+1)e^{-\frac{1}{2}l(l+1)},
\end{equation}
\noindent where we have taken a unit width to define the mother
wavelet. For $l \gg 1$ both functions approach the MHW on the
plane. We have represented the harmonic coefficients $\psi_1(lR),
\psi_2 (lR)$ for different values of $R=0.2\times 2^j, j=-2,-1,0,1,2$,
as well as the profile of the wavelets on real space for the same
cases, in Figure~\ref{fig:fig1}. The differences between wavelets
$\psi_1$ and $\psi_2$ is shown in Figure~\ref{fig:fig2}.

Next, in order to study the concentration in wavelet space we will
consider a spot with spherical symmetry placed on the north pole,
i. e. it is defined by a function $f(\theta)$. In this case, we get
for the wavelet coefficients
\begin{eqnarray}
w(R, \theta ) & = & \sum_lw_l(R)\,P_l(\cos \theta ), \ \ \ 
w_l(R)\equiv f_l\,\psi (lR), \nonumber \\
f_l & \equiv & (l+\frac{1}{2}) \int_0^{\pi}d\theta\,\sin \theta P_l(\cos
\theta )f(\theta ).
\label{eq:ct}
\end{eqnarray}
\noindent Let us now consider a very simple spot defined by a top hat
$f(\theta ) = 1_{[0, \theta_o]}$. 
%
%
%
A simple way to test how the wavelets allow us to concentrate the
information in a few number of coefficients is to measure somehow the
width of the curve $w(R, \theta )$ of wavelet coefficients for the
spot. An intuitive way to do this is to define the ``energy'' as the
integral under the squared curve $w^2(R,\theta)$ and to see which is
the radius $\theta_e$ that contains a given fraction of the total
energy. The smaller $\theta_e$ is, the more concentrated the
coefficients are.

\begin{figure*}
\begin{center}
\includegraphics[width=0.9\textwidth]{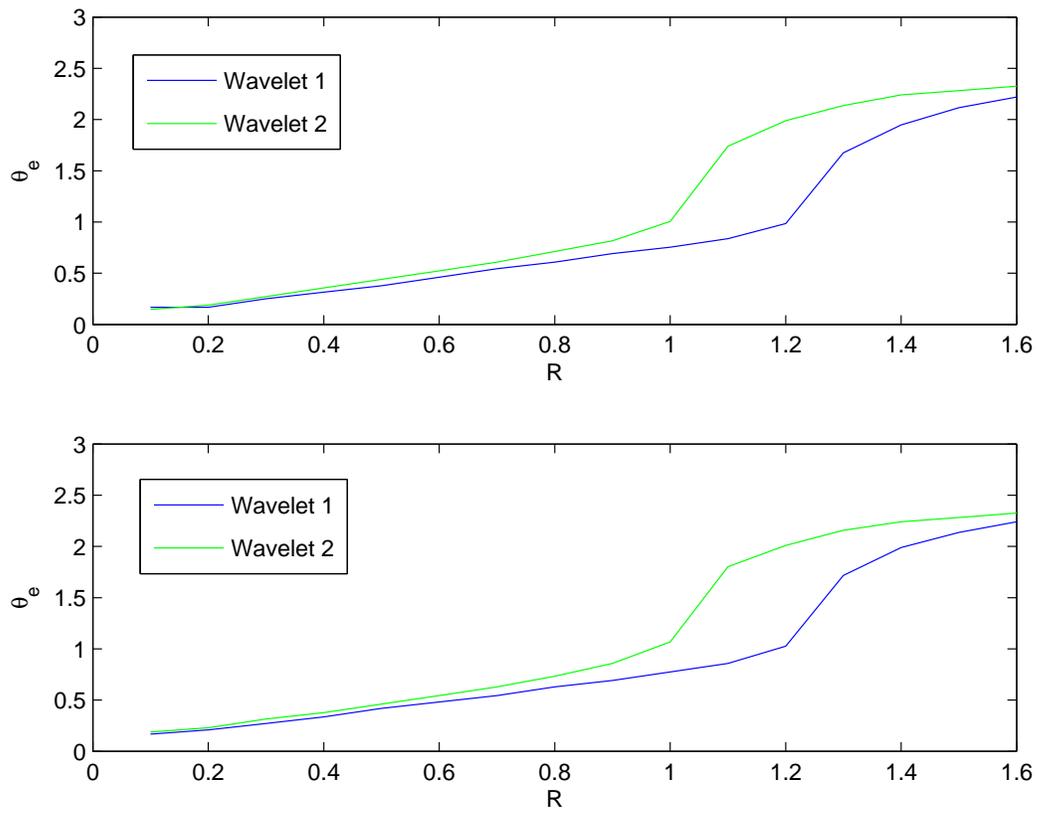} \caption{{\small Width
of the wavelet coefficient curves $w(R,\theta)$ for the wavelets
$\psi_1$ and $\psi_2$ applied to a simple top hat spot of size
$\theta_0=0.2$ rad placed on the north pole and different values of
$R$ (top panel) and for a Gaussian spot of size $\theta_0=0.2$ rad
placed on the north pole and different values of $R$ (lower panel).
The width of the curve is defined as the radius $\theta_e$ that
contains $68\%$ of the energy of the curve.
\label{fig:fig3}}}
\end{center}
\end{figure*}

We have performed numerical simulations with a simple toy model to see
which of the two generalizations of the MHW, $\psi_1$ or $\psi_2$,
concentrates more the coefficients. We have placed a top hat spot of
size $\theta_0=0.2$ rad in the North Pole and we have filtered it with
the wavelets $\psi_1$ and $\psi_2$ using different scales ranging from
$R=0.1$ to $R=1.6$ rad. The results are shown in the upper panel of 
Figure~\ref{fig:fig3}. We have ploted the radius $\theta_e(R)$ such
that $68\%$ of the energy is inside the circle of radius $\theta_e(R)$
as a function of the dilation scale $R$. As can be seen, the wavelet
$\psi_1$ concentrates more the coefficients, that is, produces smaller
values of $\theta_e$.

Now we repeat the same process but using a Gaussian spot instead of a
top hat. The Gaussian spot is given by $f(\theta ) =
\mathrm{exp}(-\theta^2/2 \theta_0^2)$. This case is interesting since
most of the detectors that operate in microwave Astronomy experiments
have approximately Gaussian response. The lower panel of
Figure~\ref{fig:fig3} show the results, that are very similar to the
top hat case. Again, $\psi_1$ concentrates more the coefficients.
We have tested the resuts for different levels of concentration and we
have found that up to $75\%$ the wavelet $\psi_1$ concentrates more
the coefficients. Above $75\%$ this is still true for large values of
the dilation $R$, while for small $R$ $\psi_2$ concentrates more the
coefficients. This is due to the fact that $\psi_1$ produces more
small oscilations in the tail of the curve $w(R, \theta )$ for small
$R$.

%
%
%

\section{Conclusions}

We have developed a constructive wavelet approach on the sphere
without any projection from the plane.  It is a continuous transform
that allows the analysis and synthesis of any function defined on the
sphere and incorporates the concepts of translation and dilation as
generalizations of the elementary ones defined on the plane. It is a
compensated filter that conserves the energy of any function.  We have
considered some natural generalizations of the plane Mexican hat
wavelet and we have applied them to the detection of a big spot. The
conclusion is that one of the wavelets ($\psi_1$) concentrates more
the information than the other one.

\section{Acknowledgements}

We acknowledge partial financial support from the Spanish Ministry of
Education (MEC) under project ESP2004--07067--C03--01. MLC acknowledge
a FPI fellowship of the Spanish Ministry of Education and Science
(MEC).  DH acknowledges the Spanish MEC for a ``Juan de la Cierva''
postdoctoral fellowship.

\newpage

\end{document}